# Localization of Energetic Frustration in Proteins

A. Brenda Guzovsky, Nicholas P. Schafer,

Peter G. Wolynes and Diego U. Ferreiro

Authors:

A. Brenda Guzovsky, Diego U. Ferreiro: *Protein Physiology Lab, FCEyN-Universidad de Buenos Aires. IQUIBICEN/CONICET. Intendente Güiraldes 2160 - Ciudad Universitaria - C1428EGA, Buenos Aires, Argentina.*

Nicholas P. Schafer, Peter G Wolynes: *Center for Theoretical Biological Physics, Department of Chemistry, Department of Physics, Department of Biosciences, 6100 Main Street, Houston, TX 77005, USA.*

**Abstract**

We present a detailed heuristic method to quantify the degree of local energetic frustration manifested by protein molecules. Current applications are realized in computational experiments where a protein structure is visualized highlighting the energetic conflicts or the concordance of the local interactions in that structure. Minimally frustrated linkages highlight the stable folding core of the molecule. Sites of high local frustration, in contrast, often indicate functionally relevant regions such as binding, active or allosteric sites.

# 1. Introduction

Biomolecules are made up of diverse parts, each falling in place, where small details of how they are put together are essential for action and life. Frustration occurs when a physical system is not able to simultaneously achieve the minimum energy for each and every part of it considered in isolation [1]. Frustration can arise from geometry or from competition between the interactions of the system's elements. The Energy Landscape Theory of protein folding applies this concept to protein molecules and provides powerful tools for understanding these evolved systems [2, 3]. Proteins are rare and wonderful polymers. They perform defined physical tasks that are seldom found by chance, such as accurate folding, specific binding and powerful catalysis. These chemical activities are the outcome of protein natural history and must be, at least partly, encoded in their sequences. In order to fold robustly, proteins must satisfy a large number of local interactions simultaneously, an optimization task feasible when frustration is low [4, 5]. Other chemical activities that must be performed, related to biological function, impose further restrictions on the sequences, possibly conflicting with the necessity of self-assembly [6]. Therefore, looking for deviations from the expected structural stability upon folding can give us hints that other teleonomic goals might be at play. Locating the frustration between conflicting goals in a recurrent system leads to fundamental insights about the chances and necessities that shape the encoding of biological information [7].

Since the global structure of a folded protein chain is the outcome of the cooperation among many local interacting units, it appears possible to decompose the global energy function into parts and quantify local conflicts as modulations of the energy change. The choice of the division into parts depends on the intention of the study, and it should be kept in mind that no subunit can be considered completely isolated. One could choose to quantify frustration of whole proteins, of domains or maybe even of "foldons", i.e. minimal folding units [8], or perhaps even go further down to contacts between amino-acid residues or atoms within residues. Having a reliable way of measuring the overall free energy of a protein structure at a given level of resolution allows one to explore how the free energy varies when the sequence or the structure of the protein changes. Ten years ago, a simple heuristic method to explore these relations was presented [9, 10]. To analyze the existence of energetic conflicts in a folded protein, the energy of structural or sequence decoys is determined and compared to that of the native state. A local frustration index is defined as the Z-score of the free energy of parts of the native structure with respect to the distribution of the energies of rearranged decoys. Interactions are divided into classes as being minimally frustrated, neutral or highly frustrated according to this index. In this chapter, we will discuss the strategy to quantify local frustration, how to calculate a local frustration index and how to analyze the results in relation to protein functions.

## 2. Materials

The basic elements needed for frustration analysis are a protein structure, an energy function (or, more precisely, a solvent-averaged free energy function) and a set of decoy protein structures. These elements should agree between each other in the level of description (or coarse-graining) of the system (Fig 1). For example, if the energy function only takes into account the positions of the Cα and Cβ, a fully atomistic description of the protein structure is unnecessary. Alternatively, if a detailed quantum-mechanical energy function is desired, highly detailed structures of protein and solvent would be required. The choice of coarse-graining will depend on the specific questions being asked about particular systems. We will use as an example of the strategy the protein model implemented in the AWSEM system [11], although the basic strategy should be applicable at every level.

2.1 - Protein structure.

A physical description of the native reference structure of the protein is required. The RCSB Protein Data Bank [12] contains about 10^5 structures, where the interested researcher may find a satisfactory model. If there is no available experimental high-resolution structure, the analysis can alternatively be well performed with a good structural prediction, whose level of resolution will depend on the energy function to be used. We found that simple threading of a sequence in a known homologous structure is generally sufficient to evaluate the frustration of most proteins (see Note 1). In general, we recommend performing various 'sanity checks' of the modelled structures [13], such as a thorough visual inspection and a structural comparison with the template and homologs [14].

2.2 - Energy function.

There are several ways to quantify the 'energy' of a particular protein structure, with varying degrees of detail and success in application, mostly depending on the question being asked. To quantify local frustration, an energy function is needed that can reliably assess how funnel-like a protein landscape is. Current approaches to analyze frustration from the ground-up, such as quantum-mechanical methods of energy evaluation, are still too computationally expensive for large macromolecules and, in any event, would require solvent averaging to be meaningful. Such a level of description is probably unnecessary for most applications related with the evaluation of local energetic frustration at the multi-atom level. Evolution works on the residue level, not the atomic scale! So while there is the possibility to employ an all-atom energy function, like those routinely used to analyze the dynamics of folded proteins [15, 16], simulating completely atomistic models in their aqueous environment and with sufficient sampling to evaluate distinct conformational states still takes a lot of computing time and may not bring too many insights [17]. At the next level of coarsening, there are many potentials that assign interaction energies to groups of atoms or pseudoatoms that aim to capture most of the relevant features of the energy landscape

[18, 19] (see Note 2). A coarser representation of the protein is in principle possible (pairs, triplets or groups of residues), although no satisfactory transferable energy function is at hand today.

2.3 - Decoy set.

Determining a frustration index depends on the choice of the parts into which the protein's whole energy is partitioned. It is useful to divide the energy up in a way that is at least roughly comparable to the energetic changes that occur in proteins in its natural or molecular history. Generally, one can examine the differences in energy upon changing the protein sequence, as relevant in evolution, or the energy changes in changing the structure, as relevant to dynamical motions. Here we present these two complementary ways for localizing frustration that differ in the way the set of decoys is constructed (Fig 2):

Mutational frustration calculations are used to answer the question, "How favourable are the native residues relative to other residues that might have been found in that location?". For mutational frustration calculations, the decoy set is made by changing the identities of the interacting amino-acids in every contact-pair or group of residues, fixing the other structural parameters at the reference value (see Note 3). Configurational frustration calculations are used to answer the question, "How favourable are the native interactions between two or more residues relative to other interactions these residues might have formed in other compact structures as the molecule folds?".For configurational frustration calculations, the decoy set is made such that the energy of the contact-pair or group of residues can be measured in globally different compact structures.

## 3. Methods

The general procedure for localizing energetic frustration consists of 7 steps (Box 1)

1. Get native (reference) structure.

2. Evaluate energy of reference.
3. Generate decoys.
4. Evaluate energies of decoys.
5. Assign frustration index.
6. Visualize and analyse results.
7. Interpret results in biological context.

Points 1-3 are discussed in the previous section (see Materials). Points 2-6 are implemented with the AWSEM energy function in a web-server (frustratometer.tk) freely available for the community [10] . A stand alone version is also available for download.

3.4 - Energies of decoys

For evaluating the energy of the decoys, care should be taken that the energy distribution of decoys reflects the natural distribution under scrutiny. In the 'mutational frustration' scheme, exhaustive mutation of every pair of contacting amino-acids is made, but we must keep in mind that the probability distribution of the occurrence of amino-acids in proteins is not uniform! [20] (see Note 4). Thus, the contributions to the decoy set must be weighted accordingly in computing variances and means. Since the distribution of single amino-acids varies over various protein families, we suggest to use the distribution that the native reference protein has or the distribution in the protein family. If an atomistic energy function is used, care should be taken to relax the structure upon making the chemical perturbation, as the inclusion of bulky residues may sterically clash with other atoms, giving unrealistically large energy differences, which might be easily relaxed. One advantage of the mutational frustration index is that, in principle, this local measure of frustration also could be experimentally determined in the laboratory by combinatorial protein engineering.

In implementing the 'configurational frustration' scheme, it is usually too computationally expensive to explicitly construct a sufficient number of globally different compact structures of a protein to evaluate the frustration index rigorously (see Note 5). Yet, a shortcut may be used. In many descriptions at low level coarsening, the energy contributions in proteins are small and numerous, such that they may be well approximated with a normal distribution, and we only need to know the mean and variance of the energy distribution that a pair of residues could explore (Fig 2B). One reasonable and computationally efficient shortcut for generating decoy structures for the configurational frustration calculation is to randomly sample the structural parameters that go into determining the energies of a contact (e.g., distances and degree of burial) from the distribution in the reference structure. The energy variance in the decoys thus reflects contributions from the energies of typical molten globule conformations of the same polypeptide chain.

3.5 - Frustration index

A frustration index ($Fi$) is defined as the Z-score of the native reference energy to the decoys:

$$Fi = \frac{E0 - \overline{Eu}}{\Delta Eu} \qquad (1)$$

Where $E0$ is the native energy, $\overline{Eu}$ is the mean of the decoy energies and $\Delta Eu$ is the variance of the decoy energies. In the case of the 'mutational frustration' scheme, it is worth noting that the energy change upon pair-mutation not only comes directly from the particular interaction probed but also changes through interactions of each residue with other residues not in the pair, as those contributions may also vary upon mutation. These changes are globally assigned to the pair-contact probed. In the 'configurational frustration' scheme, the frustration index is assigned to each contact by quantifying its native energy relative to the structural decoys.

The frustration index takes on a continuous range of values. In order to decide whether the classified interactions favor robust folding of a domain or are in energetic conflict with folding, some cutoffs in the scale need to be made. We propose cutoffs that are not unique but that we have found to be useful for understanding patterns of localized frustration in many proteins. These cutoff values are physically motivated within the energy landscape theory and are based on considering the expected ground state energy of a random heteropolymer. The basic parameters that go into determining these cutoffs are the ratio of the folding and glass transition temperatures of evolved proteins ($T_f/T_g$) and the configurational entropy change upon forming a pairwise interaction [21]. Several independent approximations to the $T_f/T_g$ ratio indicate that the original estimate for the ratio of 1.6 is pretty good [22]. The entropy change for forming a single pairwise contact depends on the sequence separation between the residues and on the degree of consolidation of neighboring interactions. Ignoring these complexities, a crude upper limit for configurational entropy change is just the sum of the entropic cost for fixing individual parts, which was estimated to be ≈0.6$k_b$ per residue in the case of amino-acid based coarsening [23]. Thus, to be minimally frustrated and fold at $T_f$, a given contact cannot be too unstable and must overcome:

$$E0 > \overline{Eu} + \delta E \times \frac{T_f}{T_g} \times \sqrt{\frac{2S}{k_b}} \times 1/\sqrt{2} \qquad (2)$$

Where $\delta E$ is the energy gap between $E0$ and $Eg$, the energy of the lowest energy misfolded state, and $S$ is the configurational entropy change for forming a single pairwise contact. Note that this is not exactly the same $\delta E$ as depicted in Fig. 2a, although the two are closely related. Satisfying this equation corresponds to having a Frustration index larger than 0.78.

Conversely, a contact will be defined as highly frustrated if $E0$ is at the other end of the distribution with a local frustration index lower than −1; that is, unlike for a minimally frustrated pair, most other amino-acid pairs at that location would be more favorable for folding than the existing pair by more than one standard deviation of that distribution. Neutral interactions are defined to lie near the center of the distribution of possible energies in compact decoy states, between the highly and minimally frustrated cutoffs. Variations of these cutoffs can readily be applied and investigated, according to the variations given by the granularity and the energy function used.

3.6 - Visualizing and analyzing results

An important aspect of frustration analysis is the visualization of the quantitative results. An intuitive representation highlighting these results can be given by coloring the pairwise contacts in the three dimensional structure of the reference. Several computer visualization programs allow one to draw lines between atoms in a protein structure that can be used to represent 'contacts' in 3D space. A psychologically (but unfortunately color-blind unfriendly) coloring scheme that became standard is to use green to represent favourable minimally frustrated contacts ("go" for folding) and red to highlight unfavourable highly frustrated contacts ("stop" for folding) (Fig 3a) (see Note 6). The frustration patterns that emerge upon exploring the structures of particular systems often can spark the comprehension of the physiology of the protein under scrutiny, at least for experts on that system. An alternative representation of frustration patterns is provided by a 'contact map' in which a matrix of interactions between every residue-pair is drawn and each contact (a point in the plot) colored according to the Frustration Index (Fig 3b). Although natural proteins are clearly not one-dimensional objects [24], in order to compare the frustration results with other common sequence-based analysis tools, a 1D representation of Frustration index along a primary structure may also be of use. One way of collapsing of the local frustration information into one dimension is to count the number of contacts that fall into each frustration class around each amino-acid (Fig 3c) (see Note 7).

Once the general aspects of the local frustration distributions are grasped, global statistical analysis of the interesting features has been performed. For example, minimally frustrated interactions tend to be found crosslinking the interior of domains, while highly frustrated interactions are found in patches at the surface of these domains [9]. A quantification of these observations can be made by calculating the pair-distribution functions of the contacts, either between each other or between these and other marked regions [25]. Analysis of the local frustration patterns over many members of protein families can identify the invariant aspects of the energy distributions and lead to the understanding of structure-function relationships [26] (see Note 8).

3.7 - Interpreting results

Interpreting the results in the biological context is the most challenging and interesting part of frustration analysis. Today, this is done by humans integrating knowledge from different sources about the particular system we are interested in. Anecdotes of local frustration distribution may be very useful to complement experimental and theoretical findings. We strongly recommend that the analysis is made having a clear hypothesis in mind with specific questions and tackling the analysis with statistics and appropriate controls [27].

The overall distribution of local frustration in proteins domains may be a useful guide for analyzing specific systems. In general, about 40% of the native interactions found in natural globular domains fall in the 'minimally frustrated' class [7]. About half of the interactions can be labelled as neutral as they do not contribute distinctively to the total energy change, and around 10% of the interactions are found to be 'highly frustrated'. These are regions in which most local sequence or structural changes would lower the free energy of the system implying that these patterns have been held in place over evolutionary time as well as physiological time at the expense of other interactions, as they conflict with the robust folding of a domain. The adaptive value for a molecule to tolerate spatially localized frustration arises from the way such frustration sculpts protein dynamics for specific functions. In a monomeric protein, the alternate configurations caused by locally frustrating an otherwise largely unfrustrated structure provide specific control of the thermal motions, guiding them in useful directions. Alternatively, a site that is frustrated in a monomeric protein may become less frustrated in the larger assembly of this protein with partners, thus guiding specific association. For a detailed discussion of the basics of frustration biophysics, the reader is referred to ref [7] and to a recent review of the outstanding applications [25].

**Notes**

1. Many algorithms are available to automatically generate protein models, with varying degrees of success [28]. We have found that the Modeller suite is reliable enough for most purposes. As in most protein model-building schemes, care should be taken in the choice of template structures, the detection of remote homologs, the crucial sequence-alignment result, and the completion of gaps, loops and missing residues. None of these computational problems are today completely solved, so watch your tools!

2. To calculate energies, we use the AWSEM potential, which was effectively constructed to infer a transferable potential for protein folding [11, 18]. The forcefield treats the polymer as being composed of three atoms per amino-acid including a sidechain of a single C$\beta$ whose interactions are amino-acid type-dependent [29]. The non-bonded interactions of the C$\beta$ can fall in any of three 'wells' of pair-wise contacts - short range, long range, and water-mediated. The side chain degrees of freedom including the rotamer state have been effectively averaged over and are implicitly encoded in the C$\beta$ interaction parameters.

3. A similar analysis can be carried out by mutating single residues. In this case, the set of decoys is constructed by shuffling the identity of the single amino-acid *i*, keeping all other parameters and neighboring residues in the native state, and evaluating the total energy change upon mutation. We call the resulting ratio the "single-residue frustration index."

4. For the mutational scheme, one effective way to take into account the different amino-acid frequencies is to sample the space of mutations with a weighted probability for each amino-acid. More detailed calculations could take into account the frequency of occurrence of pairs of amino-acids or the frequencies at each position in the family, if known.

5. For the configurational decoys in the case of AWSEM coarsening, the residues $i,j$ are not only changed in identity, but also are displaced in location, randomizing the distances $r_{i,j}$ and densities $\rho_i$ of the interacting amino-acids according to the reference distribution.

6. To visualize the frustration upon the structure, we typically do not draw the neutral contacts, as the multitude of these cover over the more interesting minimally frustrated or highly frustrated contacts.

7. Care should be taken in analyzing this collapsed representation of the contacts, as their density is typically not uniform along the sequence, and can lead to distortions in the interpretation of the plots. It should be noted that the frustration index assigned to each contact is not additive with respect to the amino-acid pair, which precludes the averaging of the frustration indexes of the contacts that each residue contributes.

8. While analyzing the local frustration patterns in a protein family, care should be taken in the alignment of the sequences or structures used, the numbering schemes, indels or gaps and other bioinformatic details that often complicate and can even preclude the analysis. Serious bookkeeping may be tedious, but is essential for making robust discoveries!

**Acknowledgment**

This work was supported by the Consejo de Investigaciones Científicas y Técnicas (CONICET); the Agencia Nacional de Promoción Científica y Tecnológica [PICT2012/01647 to D.U.F.] and ECOS Sud - MINCyT n A14E04. Additional support was provided by D. R. Bullard-Welch Chair at Rice University [Grant C-0016 to P.G.W.].

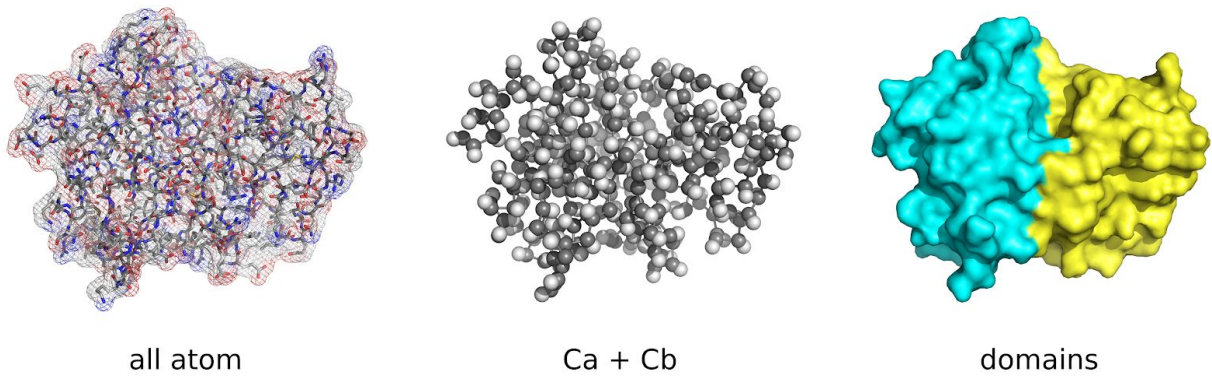

**Figure 1.** Proteins can be described using different levels of detail, depending on the specific questions for the system. These are only three of the many possible representations for the structure of *B. Licheniformis* Beta-Lactamase (pdb 4BLM). An all-atom representation (*left*) colored according to atom type. A coarse-grained representation (*middle*) showing only the Alpha (dark grey) and Beta (light grey) Carbons. A domain-level representation (*right*) showing only the two subdomains of the protein without any atomic detail.

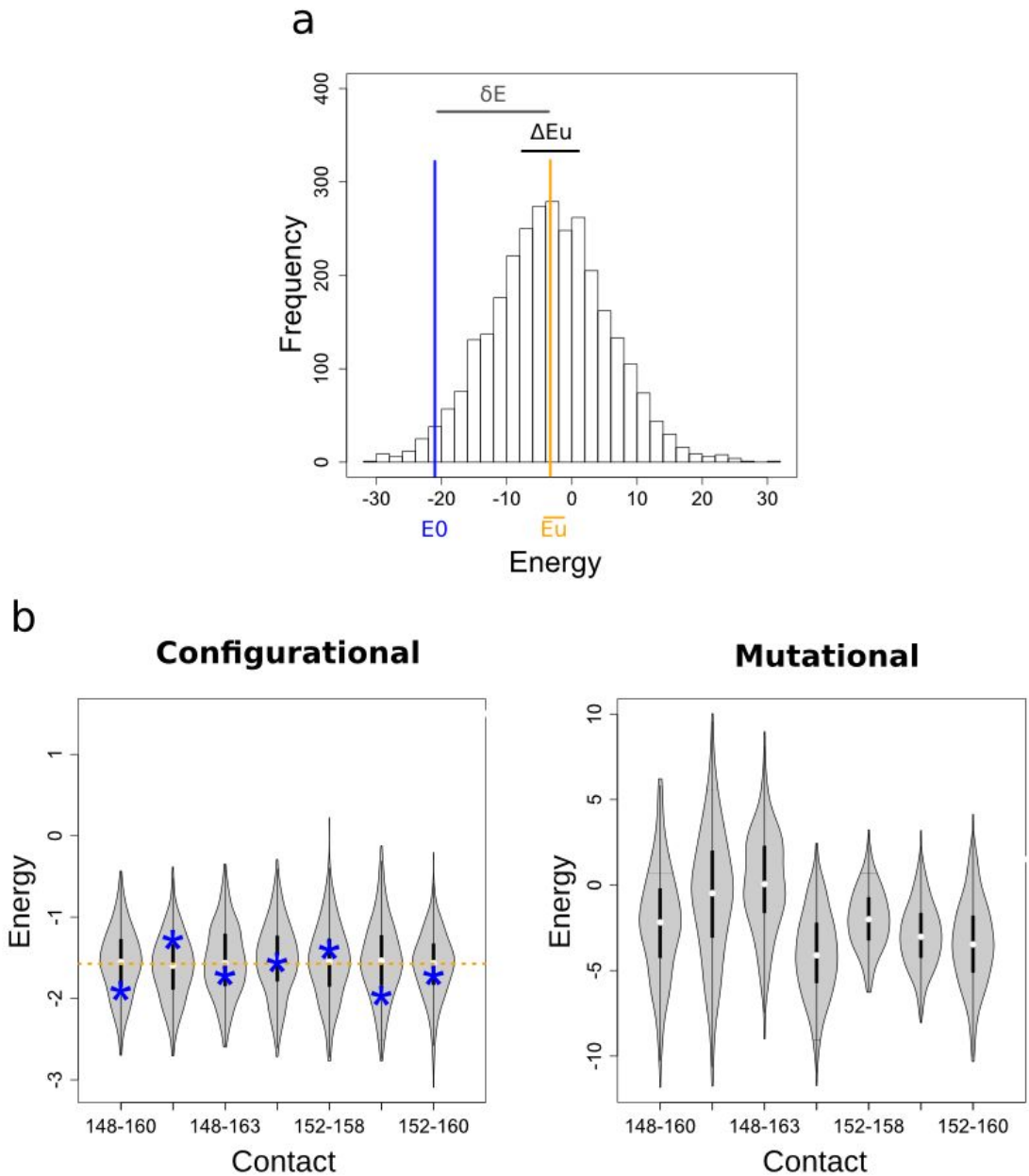

**Figure 2.** A) Energy distribution for the configurational decoys of an interaction. The energies have an approximately gaussian distribution, with a mean $\overline{Eu}$ (orange) and a standard derivation $\Delta Eu$ (black). *E0* (blue) is the native energy of the protein. For the contact to contribute to robust folding, the native and mean decoy energy should be well separated. $\delta E$ (grey) is the gap between these two energies. B)

Distribution of the decoy energies for a series of contacts. Both the configurational (l*eft*) and mutational (*right*) decoy energies are shown.

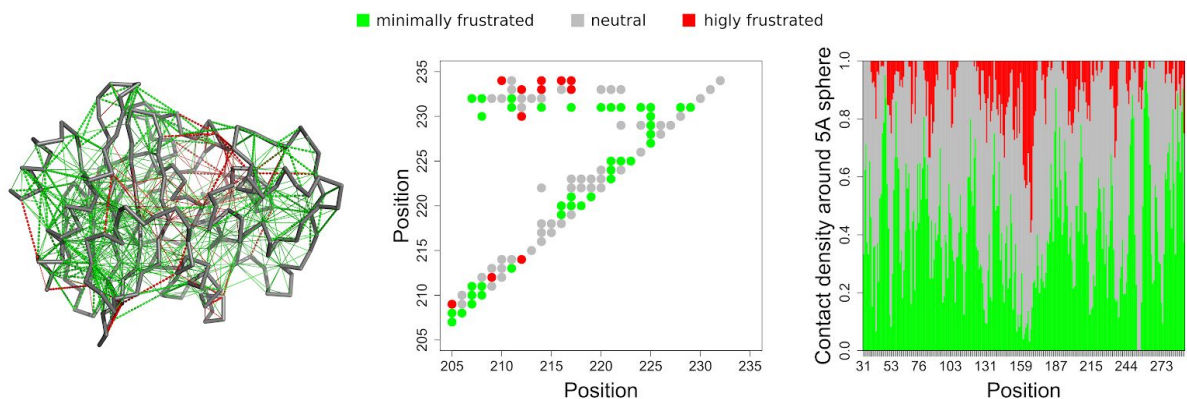

**Figure 3.** Local frustration can be visualized in 3D, 2D or 1D. In all cases, minimal frustration is represented in green, high frustration in red and neutral frustration in grey. A) Contacts are drawn on top of the 4BLM structure and colored according to their mutational frustration index. Neutral contacts are not shown in order to better enable the analysis of minimally and highly frustrated regions. B) A portion of the contact map for 4BLM. Each dot represents a contact, which is colored according to its mutational frustration index. Contacts near the diagonal are short-range, while those farther away from the diagonal make up elements of the secondary and tertiary structure. C) The density of contacts of each frustration type, according to the mutational frustration index, around 5Å of every residue. This is one way of representing the frustration index on a 1D sequence.